\documentclass[9pt,twocolumn,twoside]{opticajnl}
\journal{opticajournal} % use for journal or Optica Open submissions
\usepackage{amsmath}
\usepackage{lineno}

\usepackage[justification=justified,singlelinecheck=false]{caption}
\setboolean{shortarticle}{true}
\title{Temporal compressive edge imaging enabled by a lensless diffuser camera}

\author[1]{Ze Zheng}
\author[1,*]{Baolei Liu}
\author[1]{Jiaqi Song}
\author[2]{Lei Ding}
\author[1]{Xiaolan Zhong}
\author[3]{David McGloin}
\author[1,4]{Fan Wang}

\affil[1]{School of Physics, Beihang University, Beijing, 100191, China}
\affil[2]{School of Biomedical Engineering, Faculty of Engineering \& IT, University of Technology Sydney, NSW 2007, Australia}
\affil[3]{School of Natural and Computing Science, University of Aberdeen, King's College, Aberdeen, AB24 3FX, United Kingdom}
\affil[4]{fanwang@buaa.edu.cn}
\affil[*]{liubaolei@buaa.edu.cn}

\begin{abstract}
Lensless imagers based on diffusers or encoding masks enable high-dimensional imaging from a single shot measurement and have been applied in various applications. However, to further extract image information such as edge detection, conventional post-processing filtering operations are needed after the reconstruction of the original object images in the diffuser imaging systems. Here, we present the concept of a temporal compressive edge detection method based on a lensless diffuser camera, which can directly recover a time sequence of edge images of a moving object from a single-shot measurement, without further post-processing steps. Our approach provides higher image quality during edge detection, compared with the conventional post-processing method. We demonstrate the effectiveness of this approach by both numerical simulation and experiments. The proof-of-concept approach can be further developed with other image post-process operations or versatile computer vision assignments toward task-oriented intelligent lensless imaging systems.
\end{abstract}

\setboolean{displaycopyright}{false} % Do not include copyright or licensing information in submission.

\begin{document}

\maketitle
%\section{Introduction}
Lenses have played an essential role in optical imaging systems over the past few centuries. Traditional imaging systems designed to obtain high-dimensional images are often bulky and expensive, such as multi-shot and scanning methods for imaging multispectral or three-dimensional (3D) objects\cite{ref1}. The recent rise of the concept of lensless ‘diffuser camera’ is driving the development of miniaturized and low cost cameras\cite{ref2}. By encoding the high-dimensional information of objects, the diffuser-assisted lensless imaging systems are able to recover multiple-dimensional images such as 3D depth imaging\cite{ref3}, multispectral/ hyperspectral imaging \cite{ref4,ref5,ref6}, 3D fluorescence imaging\cite{ref7}, compressive temporal imaging\cite{ref8,ref9}, full-Stokes polarization imaging\cite{ref10}, multi-modality edge enhancement imaging\cite{ref11} and so on. Generally, the starting point for such devices is the characterization of the diffuser, such as calibrating the point spread functions (PSFs) of a ground glass diffuser or phase mask. Then a two-dimensional (2D) image is captured by putting the diffuser, instead of a lens, between an object and a 2D sensor. Finally, inverse algorithms are implemented to recover high-dimensional images. Recent advances in diffuser cameras have shown great promise in both applications and adding new imaging functionality. For example, in vivo lensless 3D microscopy has been achieved by using a specially designed phase mask\cite{ref12} and multi-dimensional imaging, including spatial, spectral, and polarization dimensions, can be encoded by a liquid crystal metasurface diffuser\cite{ref13}. In addition, programmable diffusers, such as reconfigurable particle assembly masks\cite{ref14}, show potential for developing tunable diffuser cameras.

After the images of an object are obtained, post-image processing is critical for acquiring additional useful information\cite{ref15}. For example, edge detection has been widely used in advanced driving assistance systems and geographic environment monitoring\cite{ref16,ref18}. Since the results of the diffuser camera are reconstructed from the raw data, a fusion of the traditional inverse algorithms used in diffuser cameras and image post-processing algorithms could generate the post-processed images directly from the raw data, without recovering the original images of the objects. Such an implicit image processing method, which is involved in computational optical imaging systems, would offer advantages over the post-processing method, offering higher signal-to-noise ratios (SNRs) of edge detection in the case of computational ghost imaging\cite{ref19,ref20,ref21}.

In this work, we propose a temporal compressive edge detection method based on a lensless diffuser camera (Diffuser-eCam), which can directly recover a sequence of edge images of an object at different time points from a single-shot measurement, without further post-processing steps. By only modifying the forward model matrix of the diffuser camera in the reconstruction, rather than modifying any steps in the experiment, we can incorporate the image post-processing steps into the reconstruction algorithm. To demonstrate this, we introduce an edge-detection filter into the model matrix, which serves as a spatial filtering operation, allowing the inverse algorithms to directly reconstruct the edge images of an object, without recovering its original image. With both numeral simulations and experimental results, we show this method can improve the peak signal-to-noise ratios (PSNR) and information entropy (IE) of the resulting images more than a conventional post-processing method using the same filter. Furthermore, we show how to recover multi-frame edge images of a moving object from a single-frame 2D measurement, by exploring the rolling-shutter mode of the 2D sensor. Other image post-processing operations can be studied in diffuser cameras or other computational imaging systems, to bypass the post-processing steps and enhance their resulting image quality.

\begin{figure}[ht]
\centering
\includegraphics[width=\linewidth]{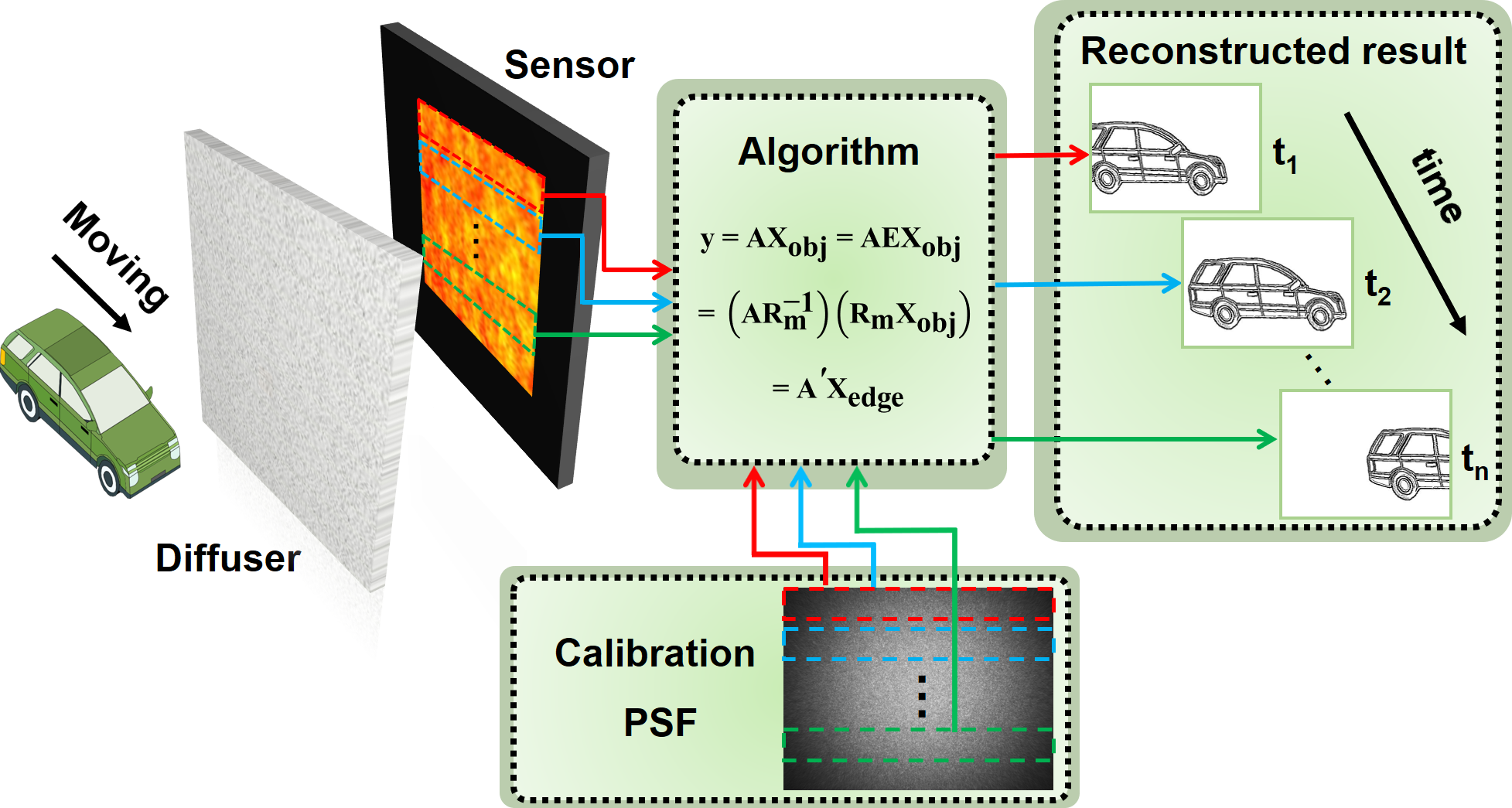}
\caption{Principle of the lensless diffuser camera for temporal compressive edge detection (i.e. Diffuser-eCam). The system encodes the temporal moving object in a 2D image on the chip CMOS sensor. The sensor works in a rolling shutter mode which different rows of the CMOS sensor begin to be exposed at different times. For calibration, a 100 $\mu$m pinhole is placed at the object plane for the capture of the system’s PSF, as shown at the bottom. By only modifying the forward model matrix A in the reconstruction, the edge images of the object can be directly reconstructed by inverse algorithms. The multi-frame edge images of the moving object are reconstructed from different rows of the measurement and corresponding rows of the PSF image.}
\label{fig:1}
\end{figure}

%\section{PRINCIPLE}
A diffuser camera is usually considered as a lensless imaging system that uses a random diffuser, instead of the traditional lenses to modulate the light field, as shown in Fig. 1. The camera sensor no longer receives an image of the object, but a blurry speckle pattern. In this work, we assume a 2D object is located within the region of the angular memory effect of a diffuser, in which the PSFs corresponding to different point sources do not change their shape and only translate in the image plane\cite{ref22}. Another assumption is that all the point sources within the object are incoherent with each other. Then, the 2D measurement \emph{I} can be expressed as the convolution of the object \emph{O} and the system’s PSF \emph{P} :

\begin{equation}
I(x,y) = O(x,y) \star P(x,y),
\label{eq:refname1}
\end{equation}
where $\star$ represents convolution operation. For convenience, we can also express it in the form of matrix multiplication:

\begin{equation}
y = AX_{obj},
\label{eq:refname1}
\end{equation}
where \emph{y} is the column vector composed of intensity values of different pixels of \emph{I}; \emph{$X_{obj}$} is the column vector composed of the each intensity value of \emph{O}; \emph{A} is the forward model matrix or calibration matrix that is related to the light modulation of the diffuser, and each of its columns can be obtained by translating and stretching the diffuser’s PSF. By using inverse algorithms, the original image \emph{$X_{obj}$} can be reconstructed.

To extract the edge features of the object, conventional methods require post-processing operations on the original image \emph{$X_{obj}$}, which needs to be reconstructed in advance. Here we choose an edge-detection operator \emph{R}\cite{ref20} to demonstrate this effect:
\begin{equation}
R = 
\begin{pmatrix}
0 & -1 & 0 \\
-1 & 0 & 1 \\
0 & 1 & 0
\end{pmatrix}.
\label{eq:1}
\end{equation}

The desired result \emph{$X_{edge}$} can be yielded by convoluting  \emph{$X_{obj}$} and the edge-detection operator \emph{R}. It can also be expressed as the form of matrix multiplication:
\begin{equation}
X_{edge} = R \star O(x,y) = R_{m}X_{obj},
\label{eq:2}
\end{equation}
where \emph{$R_{m}$} is the corresponding matrix form of the edge-detection operator R. We can use the identity matrix \emph{$E=R_{m} R_{m}^{-1}$} to modify Eq. (2):
\begin{equation}
y = AX_{obj} = AEX_{obj} = (AR_{m}^{-1})(R_{m}X_{obj}) = A^{'}X_{edge},
\label{eq:3}
\end{equation}
where \emph{$A^{'} = A{R_{m}}^{-1}$} is the modified forward model matrix that can be generated before the reconstruction starts. Then, the compressed sensing algorithm, such as compressive sensing with total variation regularization\cite{ref23} and the TwIST algorithm\cite{ref24}, can directly extract the object’s edge \emph{$X_{edge}$}:
\begin{equation}
\hat{X}_{edge} = arg\min_{X_{edge} \geq 0} {\lVert y - A^{'}X_{edge} \rVert}_2^2 - \tau {\lVert \Psi X_{edge} \rVert}_1.
% \hat{X}_{edge} = argmin\limits_{X_{edge} \geq 0} {\lVert y - A^{'}X_{edge} \rVert}_2^2 - \tau {\lVert \Psi X_{edge} \rVert}_1
\label{eq:4}
\end{equation}
here, $\Psi$ is the linear change matrix that maps \emph{$X_{edge}$} to the domain with sparse representation, and \emph{$\tau$} is a regularization parameter that tunes the sparsity of the scene.

%\section{RESULT}

To demonstrate the proposed method, we first conduct the numerical simulation shown in Fig. 2. The original images, which are the grayscale versions of the images of ‘hand’, ‘foggy road’, and ‘tumor tissue’ from MATLAB’s image data, are shown in the left column of Fig. 2. The PSF used is experimentally measured from a ground glass diffuser (GCL-201101, DHC). The results from the Diffuser-eCam and conventional post-processing method at different sampling rates are shown in Fig. 2 and indicated with ‘Diffuser-eCam’ and ‘post-processing’, respectively. The edge images of ‘Diffuser-eCam’ are directly reconstructed from Eq. (6). The ‘post-processing’ method requires the reconstruction of the image of the objects first (Eq. (2)), and then the convolution of the object images and the edge-detection operator according to Eq. (4). 

The image quality of the results, for both methods, increases with the sampling rate. Note that the resulting images show a kind of relief effect, which is similar to that produced by differential interference contrast\cite{ref25}, due to the asymmetry of the operator used in Eq. (3). It is evident that the results of “Diffuser-eCam” have higher contrast and more detail than those from the “post-processing” method. For example, the edges of the trees that are obscured by fog, as shown in Fig. 2(b) are recognizable by “Diffuser-eCam” and barely visible by the “post-processing” method, while the edges of roads are clearly shown in both the two methods. For the results of ‘hand’ and ‘tumor tissue’, the results of Diffuser-eCam are more prominent and have higher image contrast than the post-processing method. The reason for the phenomenon can be explained that objects' edges have much higher sparsity than the original objects' images\cite{ref21}. It is also noted that the quality of edge detection, for the majority of real objects, is improved if the object’s edge can be reconstructed directly, compared with the traditional procedure that needs to reconstruct the original object image in advance\cite{ref19,ref20,ref21}. 

\begin{figure}[ht]
\centering
\includegraphics[width=\linewidth]{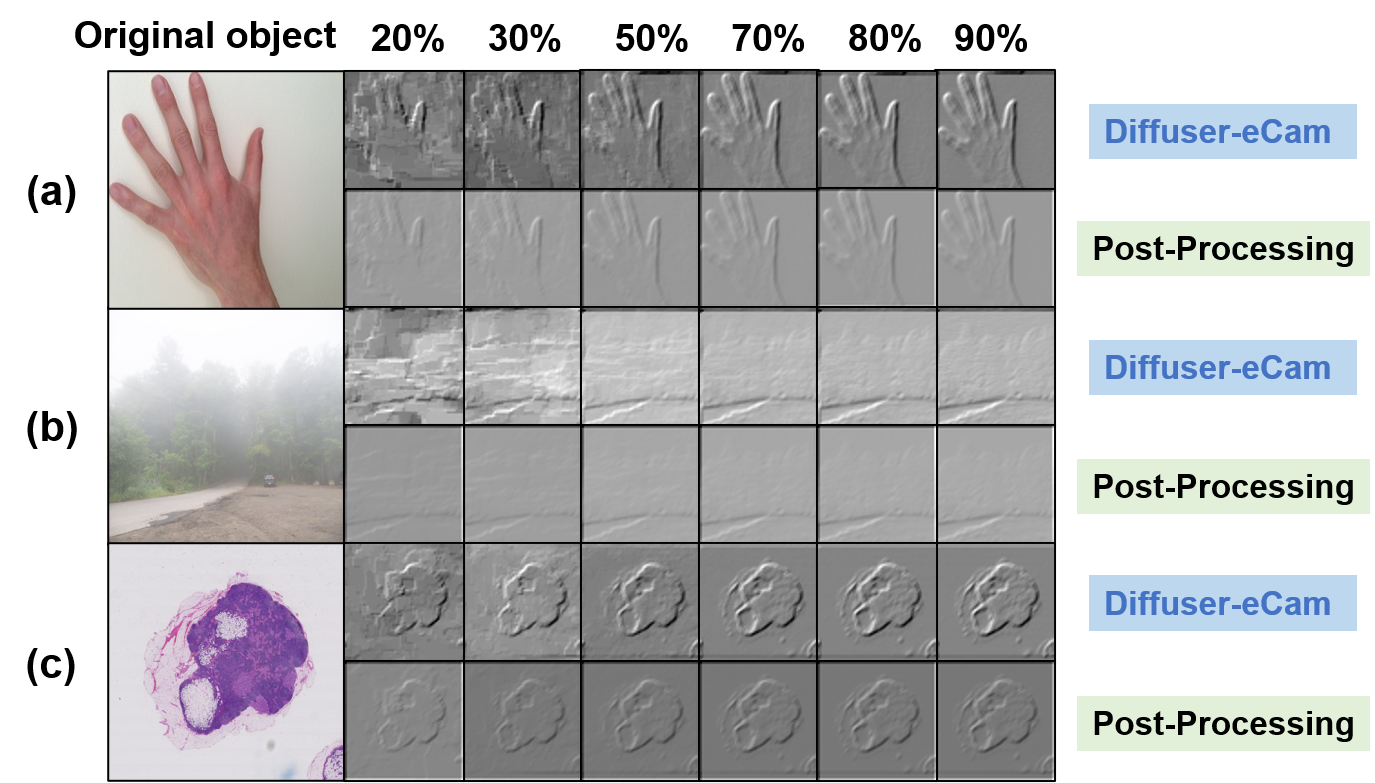}
\caption{Simulated results of edge detection by using Diffuser-eCam and conventional post-processing methods with sampling rates of 20\%, 30\%, 50\%, 70\%, and 90\%, respectively. The original objects are the images of (a) ‘hand’, (b) ‘foggy road’, and (c) ‘tumor tissue’.}
\label{fig:false-color}
\end{figure}

Then we conducted experiments to verify the proposed method. The light source is a monochromatic LED (GCI-060401, DHC), which is diffused by the ground glass and collimated by a convex lens (\emph{f} = 50 mm). The objects are 3D-printed transmitted masks (about 3 cm×3 cm) with four designs: the letter ‘T’, three stripes, the up arrow, and the U-turn arrow. The distance between the objects and the LED was 15 cm. The light transmitted through the objects is scattered by a diffuser (GCL-201101, DfHC). The distance between the object and the diffuser was 28 cm. A 4 mm diameter aperture (GCT-212621-6, DHC) was placed close to the diffuser to block the background light. The light field after the diffuser was measured by a CMOS sensor (daA2500-14um, Basler), which was placed 1 cm behind the diffuser. The exposure time was set to 4 ms.

We placed a 100 $\mu$m pinhole at the object plane to capture the system’s PSF image, with an exposure time of 500 ms. The PSF only needs to be calibrated once as the objects are within the range of the optical angular memory effect. The resulting images are evaluated by PSNR and IE\cite{gonzalez2009digital}. PSNR is calculated based on mean squared error (MSE), which measures the pixel difference between the reference image \emph{K(x,y)} and the reconstructed image \emph{I(x,y)}. A higher PSNR value indicates a smaller difference in pixels’ value between the reconstructed and reference edge image. PSNR is defined as:
\begin{equation}
PSNR = 10\log \frac{MAX^{2}}{MSE},
\label{eq:refname1}
\end{equation}
where MAX represents the maximum value of image pixels  (255 for 8-bit images). The reference edge image is post-processed with the object image which we capture in the same position using a commercial sensor (IMX586, SONY). The MSE is defined as:
\begin{equation}
MSE = \frac{1}{mn} \sum\limits_{i=0}^{m-1} \sum\limits_{j=0}^{n-1} \left[I(i,j) - K(i,j)\right]^{2}.
\label{eq:refname1}
\end{equation}

The IE is used to measure the information quantity or uncertainty of an image, to quantify the complexity and information content of edge images. The edge of an image typically represents gradient changes of pixel values and structural information. Therefore, the IE of edge images can reflect the richness or complexity of edges. The IE is defined as: 
\begin{equation}
IE = - \sum\limits_{i=1}^{n} P_{i}\log_2 P_{i},
\label{eq:refname1}
\end{equation}
where \emph{$P_1,P_2,… ,P_n$} represents the probability distribution of different pixel values in the normalized image.

\begin{figure}[ht]
\centering
\includegraphics[width=\linewidth]{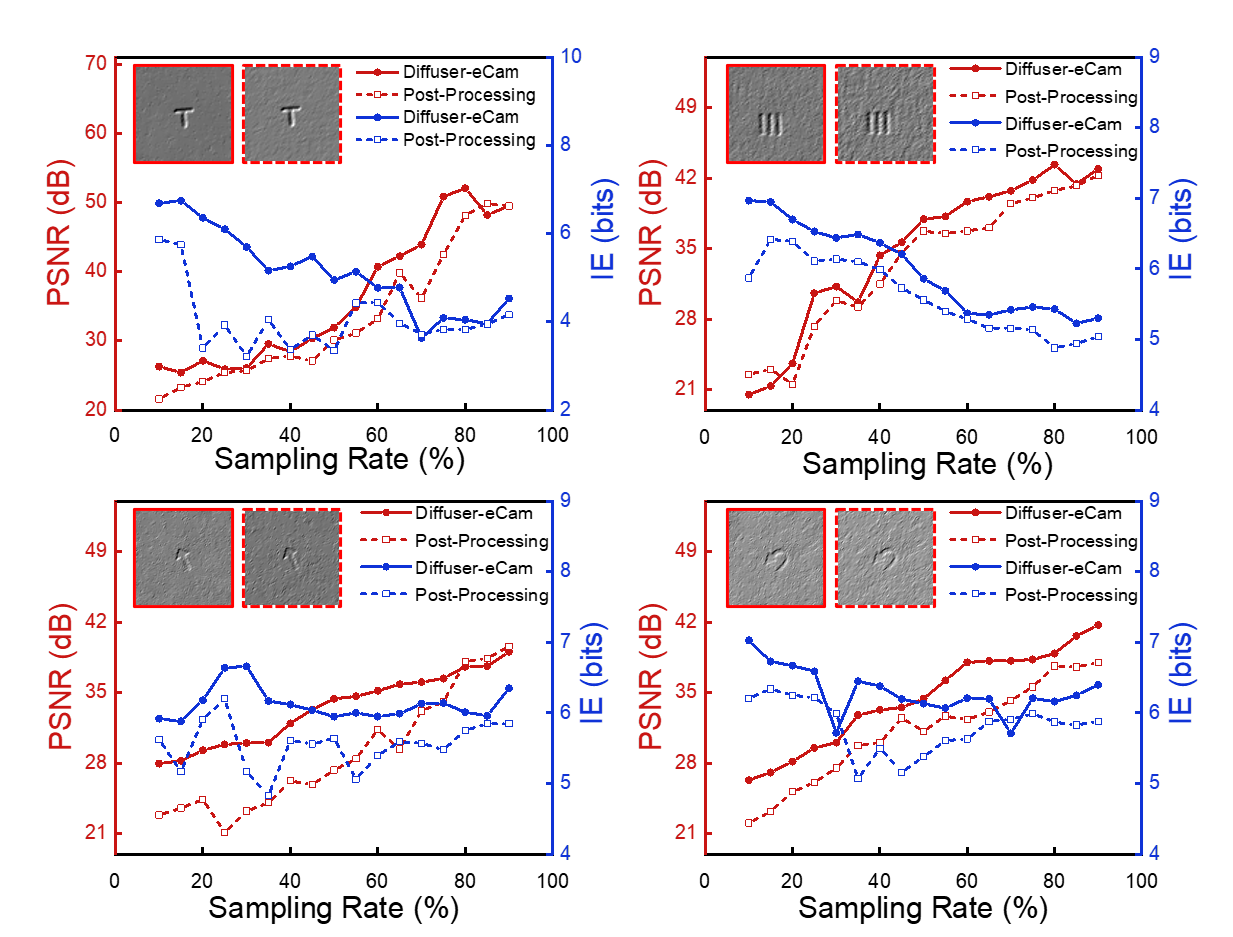}
\caption{Experimental results and comparison of Diffuser-eCam and conventional post-processing method. The example resulting images of four different objects are shown in the upper left corner of each subfigure, with a sampling rate of 30\%. The image quality of the results, at different sampling rates, is characterized by using PSNR and IE. The red lines represent the PSNR, and the blue lines represent the IE values. The solid and dashed lines are the calculated values from Diffuser-eCam and the conventional post-processing method, respectively.}
\label{fig:false-color}
\end{figure}

Figure 3 compares the post-processing method and Diffuser-eCam for three different objects. With a sampling rate of 30\%, the results from Diffuser-eCam have less noise and smoother backgrounds than the post-processing method, as shown in the inset figures. The PSNRs of images from the two methods, at different sampling rates, are shown as the red solid and dashed lines, respectively. Generally, the PSNRs for different objects increase as the sampling rate increases. Compared with the post-processing method (red dashed lines), the resulting images of Diffuser-eCam have higher PSNRs and less noise (red solid lines). The corresponding values of information entropy are shown as the blue solid and dashed lines. 

As shown in Fig. 3, the solid lines, whether red or blue, are almost above the dashed lines. In other words, the PSNRs and IE values of the results from Diffuser-eCam are mostly higher than the conventional post-process results. The edge is usually the place where the pixel value changes most obviously in grayscale. Better edge quality results in a higher IE value and a higher PSNR value that indicates lower noise. It implies that the contribution to the growth of the PSNR and the IE is mostly generated from the sharper edge rather than the more complex background. Thus, it can be concluded that Diffuser-eCam has improved the image quality of the objects’ edge, compared with the conventional post-processing method.  

\begin{figure}[ht]
\centering
\includegraphics[width=\linewidth]{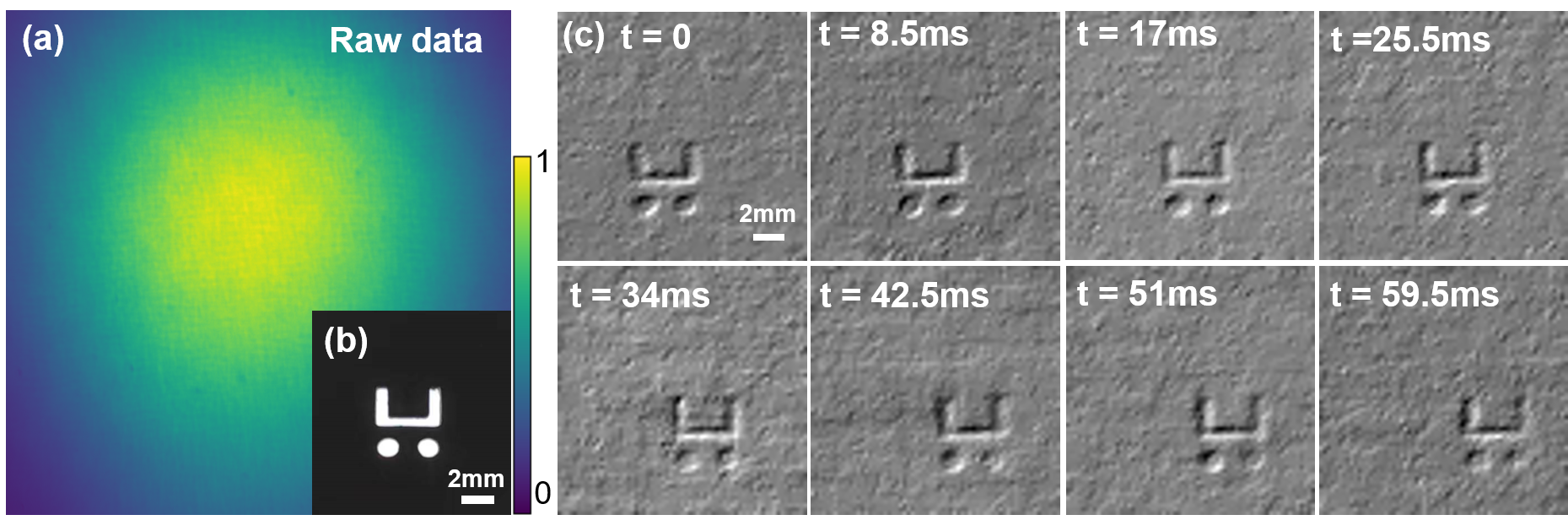}
\caption{Demonstration of reconstructing multi-frame edge images at different times of a moving object, from a single 2D measurement. (a) The 2D raw measurement is captured by the CMOS sensor with the rolling shutter mode. (b) The photo of the 3D-printed ‘car’ object, which would be moving laterally in the experiment. The object is about 3 mm×3 mm and the white parts are transmitting. (c) The reconstructed edge images are at different time points. The sequential images are reconstructed from different rows of (a), which are exposed at different times in the rolling shutter mode.}
\label{fig:false-color}
\end{figure}

One advantage of diffuser cameras is the ability to achieve compressive temporal imaging that encodes a temporal scene into a single-shot measurement by using the rolling-shutter mode of the sensor, in which the pixels in each row of the sensor are sequentially exposed at different times. To demonstrate the reconstruction of multi-frame edge images of a moving object from a single-shot 2D measured speckle, as shown in Fig. 4, we make use of a 3D-printed ‘car’ as the object, as shown in Fig. 4(b). The ‘car’ is mounted on a translation stage (GCM-083904M, DHC) and is moved parallel to the diffuser camera in the experiment. The raw data measurement is shown in Fig. 4(a), with an exposure time of 4 ms. Since different rows of the measurement correspond to different times, the resulting image at a certain time can be reconstructed from the corresponding rows of the measurement. The multi-frame edge images of the moving ‘car’ are shown in Fig. 4(c), corresponding to different times (T=0,8.5ms,\dots,59.5ms). We can clearly observe that the ‘car’ is moving from the left to the right. The ‘car’ moved approximately 5mm during the exposure time.

%\section{CONCLUTION}
In conclusion, we have proposed and demonstrated temporal compressive edge imaging based on a lensless diffuser camera from a single-shot measurement. The diffuser camera encodes a temporal scene into a 2D image on the sensor. By only modifying the forward model matrix in inverse algorithms, the edge images of an object corresponding to different times can be directly reconstructed from different rows of the 2D raw measurement. The proposed method can achieve higher image quality, compared with the conventional post-processing method that convolves the retrieved object image with an edge-detection operator. Thus, Diffuser-eCam shows not only a new dimension of diffuser cameras for edge detection but also a higher resulting image quality than the conventional post-processing method.

One major advantage of the proposed method is that it does not require any change to the experimental setup or multiple measurements, but only modifying the forward model matrix during the reconstruction process. Therefore our method would inspire further developments of diffuser cameras with the realization of other digital image processing tasks, such as high/low pass filtering, deconvolution, denoising, and so on. Further complex vision tasks, such as face detection and fingerprint identification, can be also considered. The simple hardware system also makes the diffuser camera easily add new imaging dimensions, such as wavelength and polarization, into the imaging systems. Our scheme provides a new way to realize edge detection based on a lensless diffuser camera. We, therefore, anticipate that this work will open opportunities for developing smart lensless imaging systems with versatile vision tasks. 

%\section{backmatter}

\begin{backmatter}
\bmsection{Funding} 
China Postdoctoral Science Foundation (2022M720347); National Natural Science Foundation of China (62075004, 11804018, 62275010); Beijing Municipal Natural Science Foundation (4212051, 1232027); the International Postdoctoral Exchange Fellowship Program (YJ20220241); the Fundamental Research Funds for the Central Universities.

% \bmsection{Acknowledgments} 
% 

\bmsection{Disclosures} 
The authors declare no conflicts of interest.

\bmsection{Data availability} 
Data underlying the results presented in this paper are not publicly available at this time but may be obtained from the authors upon reasonable request.

\end{backmatter}

\bigskip
% \noindent Add citations manually or use BibTeX. See \cite{ref1}.

% Bibliography
\bibliography{bibliography}

% Full bibliography added automatically for Optics Letters submissions; the following line will simply be ignored if submitting to other journals.
% Note that this extra page will not count against page length
\bibliographyfullrefs{bibliography_full_name}

%Manual citation list
%\begin{thebibliography}{1}
%\bibitem{Zhang:14}
%Y.~Zhang, S.~Qiao, L.~Sun, Q.~W. Shi, W.~Huang, %L.~Li, and Z.~Yang,
 % \enquote{Photoinduced active terahertz metamaterials with nanostructured
  %vanadium dioxide film deposited by sol-gel method,} Opt. Express \textbf{22},
  %11070--11078 (2014).
%\end{thebibliography}

% Please include bios and photos of all authors for aop articles

\end{document}